# ASFAP Working Groups Activity Summary: Biophysics, Light Sources, Atomic and Molecular Physics, Condensed Matter and Materials Physics, and Earth Sciences

Sonia Haddad[a], Gihan Kamel[b*], Lalla Btissam Drissi[c], Samuel Chigome[d,]

[a]*Laboratoire de Physique de la Mati`ere Condens´ee, Facult´e des Sciences de Tunis, Universit´e Tunis El Manar, Campus Universitaire 1060 Tunis, Tunisia*
[b]*Department of Physics, Faculty of Science, Helwan University, Egypt, and SESAME Synchrotron Light Source, Jordan*
[c]*Faculty of Sciences, Mohammed V University, Rabat, Morocco*
[d]*Botswana Institute for Technology Research and Innovation, Gaborone, Botswana*

**Abstract**

Various panel sessions were organized to highlight the activities of the African Strategy for Fundamental and Applied Physics (ASFAP) Working Groups during the second African Conference of Fundamental and Applied Physics (ACP2021) that was held in March 7–11, 2022. A joint session was devoted to highlight the activities assigned to the Light Sources, Accelerators, Biophysics, Earth Sciences, Atomic and Molecular Physics, and Condensed Matter and Materials Physics Working Groups. Major outcomes and recommendations are demonstrated and deliberated in this contribution.

## Introduction

The African Strategy for Fundamental and Applied Physics (ASFAP) aims at implementing state- of-the-art scientific and technological knowledge and development in Africa based on a concrete and tangible strategic vision in a multi-fold approach. In an attempt to create a focal point for capacity building, retention and advanced research, various specialized working groups were established to raise the awareness about the ASFAP objectives, activities, as well as, the envisaged deliverables. In this regard, six Working Groups representing Light Sources, Accelerators, Biophysics, Earth Sciences, Atomic and Molecular Physics, and Condensed Matter and Materials Physics did contribute to the ACP2021. The groups' objectives and activities were presented. In addition, several outcomes of mutual discussions highlighting the received Letters of Intent (LoIs) and the published surveys were provided.

\* Corresponding Author (Gihan Kamel)

1. **Biophysics WG**

The ASFAP Biophysics working group tackles one of the most demanded disciplines in Physics. The Working Group aims at assessing the scope of the biophysics education in Africa, extending the biophysics research database, as well as, development of biophysics curricula. The group has received seven LoIs and is expecting to receive more letters of support as a result of the subsequent planned on-line conference titled "Biophysics in Africa" which was also organized by this group in March 21-25, 2022 [3]. This conference shed the light on the different topics of biophysics, in particular; molecular biophysics, structural biology, maternal biophysics, computational biology, quantum biology, radiology, and biophotonics.

Among the various activities, the Biophysics Working Group has launched a call for Letters of Intent during the organized Biophysics Winter School (4 July 2022) [5], and also at the Structural Biology Workshop (April 25-28, 2022). The group proposed to extend the biophysics research databases and to work in liaison with other ASFAP WGs, as well as with other related international organizations such as the International Union of Pure and Applied Physics, IUPAP, and with the Biophysical Society [6]. In order to have more synergy with other groups, the Biophysics WG did propose to be merged with other ASFAP Working Groups as well namely the Light Sources and the Nuclear Physics ones.

2. **Light Sources WG**

The ASAFP Light Sources WG is mandated to highlight the demands of the wide scientific community in different disciplines, and to bring the important role that light sources including synchrotron radiation facilities to the front-lines. Such large-scale infrastructures are playing a key role in scientific and technological advancement of any society. Many challenges do exist in Africa such as establishing cutting-edge research infrastructures and institutions, reversing the brain-drain dramatic challenge, addressing concerns such as health, environment, water, energy, and climate change, etc. In addition, the WG is also aiming at boosting the visibility of the fundamental significance of such a competent light source facility in Africa.

In an attempt to implement such goals, the Light Sources WG launched a comprehensive survey in which physicists from different African countries and diaspora have also participated (Fig.1). According to this survey [9], there is an essential need to establish a light source in Africa, which is the only continent that is left behind without such a large-scale infrastructure thus far.

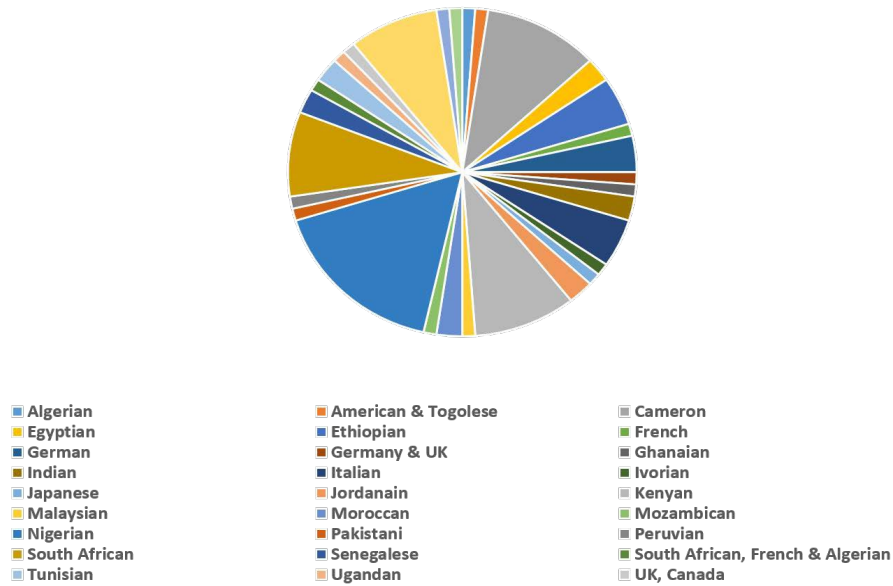

**Figure** 1: Origin of the participants to the survey launched by the Light sources working group [7, 9].

Furthermore, the survey conveyed a practical vision that the African countries should share existing infrastructures while preparing for a common platform which can then be guided by a scientific council involving African Physicists and third parties like UNESCO for instance. It is worth to note that only 13% of the participants to this questionnaire were female students and experienced researchers, which brings into the surface the necessity of minimizing the gender gap in Africa.

The survey has also underlined the challenging difficulties pausing so many African Physicists in pursuing an advanced scientific career in worldwide light sources (Fig.2). The participants have also provided some recommendations about the current educational systems in African countries. For that, they proposed to include Light sources training sessions and lectures in the African curricula of graduate students. They also suggested to involve local industry and engage policymakers and the international community in order to support the strategic vision for African targeted infrastructures and light sources facilities.

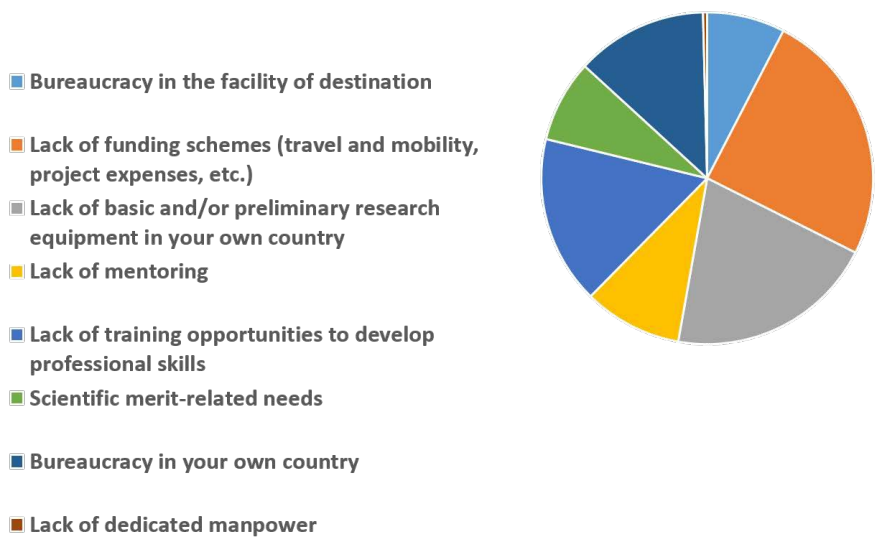

**Figure** 2: Difficulties facing scientists in Africa as has been raised by participants to the Light sources survey [7, 9].

In addition to the survey, several LoIs were submitted in favor of the establishment of the first light source in Africa. The WG is currently analyzing these received letters, and intends to advocate for more through complementary activities.

### 3. Atomic & Molecular Physics WG

The ASFAP Atomic and Molecular Physics (AMP) working group's strategy is to identify and approach similar working groups in Africa, and African diaspora in the field, focusing on quantum physics/technologies researchers, electronic structure communities like ASESMA, CASESMA, as well as, the optics groups in Cameroon. Out of the main disciplines is that the AMP WG is tasked to monitor the activities of various scopes such as Atomic structure, properties, and dynamics, Molecular, Chemical and Cluster Physics, Cold Matter; Quantum Technologies; and Astrophysics and Plasma Physics. The working group [8] has also organized an online mini-workshop on $10^{th}$ January 2022 where a hundred of participants had registered. However, due to some connectivity issues, only 30% to 40% of the participants managed to attend. Concerning the LoIs, the WG on Atomic and Molecular Physics (AMP) has received only seven LoIs among others where the AMP disciplines are not the primary field. The best strategy to overcome this drawback is to merge the AMP WG with other ASFAP working groups to boost the foreseen goals. The outcomes of the discussion session on AMP brought out the necessity to get Africa involved in the rising worldwide quantum initiatives and quantum technologies, as Africa is left behind in the nationwide investments in such technology (Fig.3).

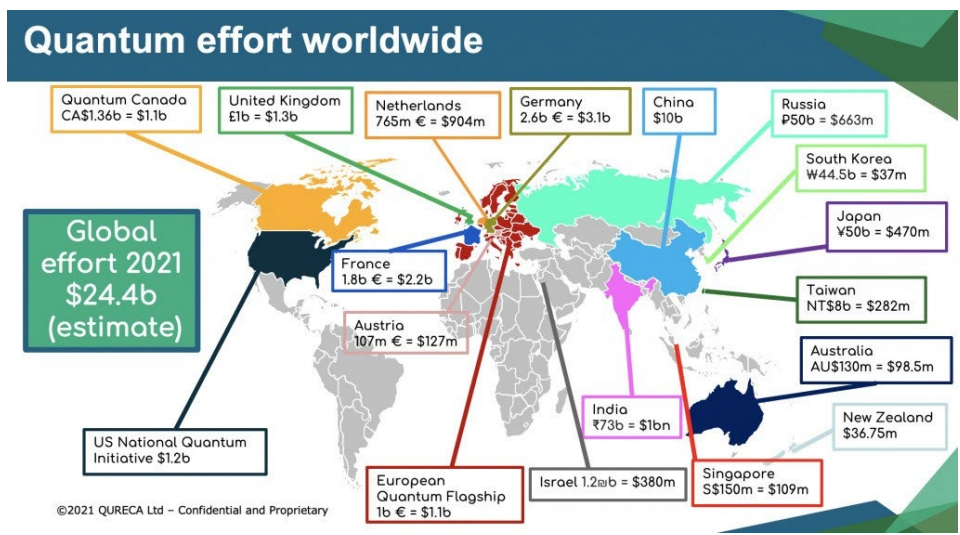

**Figure 3**: Quantum initiatives worldwide [8].

### 4. Condensed Matter & Materials Physics WG

The Condensed Matter & Materials Physics (CM) WG [10] has organized an online mini-Workshop on 15th December 2021 [11] where more than 200 participants attended with a rate of 38% of full workshop attendance. However, only 28% of female participation was observed (Fig.4). This workshop included an introduction to ASFAP, presented by members of the steering committee, as well as, some plenary talks aimed to raise the awareness about the importance of the Condensed Matter and Materials Physics in Africa developmental path [11].

The CM WG has also launched a survey to collect recommendations which can then be structured within the LoIs [12]. The survey has received, till June 28$^{th,}$ 2022, a total number of 209 responses. During the ACP 2021, a percentage of more than 55% responses has been reached with a total of 198 responses were collected. It was emphasized that the survey should be advertised during all the events held in Africa to attract more participants. The survey's participants comprise different areas in Condensed Matter Physics. Figure 5 shows the different relevant areas to the condensed matter physics as the respondents of the survey have indicated. The survey has given a clear evidence that there is a huge lack in equipment for African experimentalists and theorists as shown in Figure 6. On the other hand, the survey also indicated that there is an urgent need to improve the standing of the curricula in Master and Doctorate levels, and to establish joint inter- Africa post-graduate degrees. Another recommendation highlighted the importance that the educational courses should be extended to hands-on sessions in research laboratories and engagement of industrial sectors (Fig. 7).

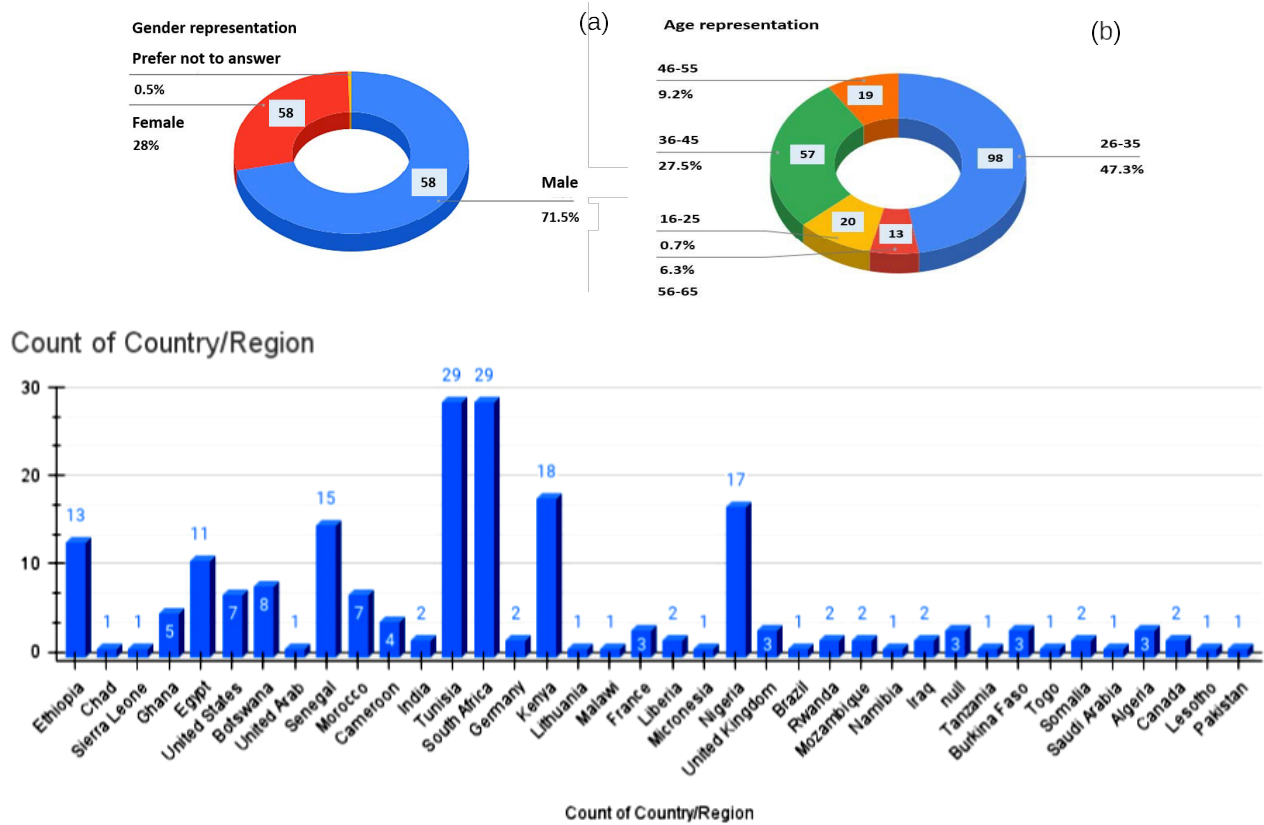

**Figure 4**: Statistics of the on-line mini-workshop organized by the Condensed Matter and Materials Physics WG: (a) The gender participation ratios. (b) The age distribution of the attendees [10] and (c) the number of participants per country.

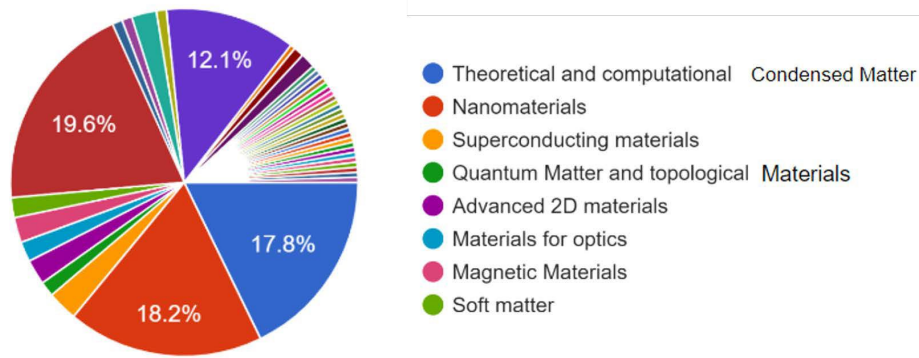

**Figure 5**: Some favorable research areas of the participants to the survey launched by the Condensed Matter and Materials Physics WG [10].

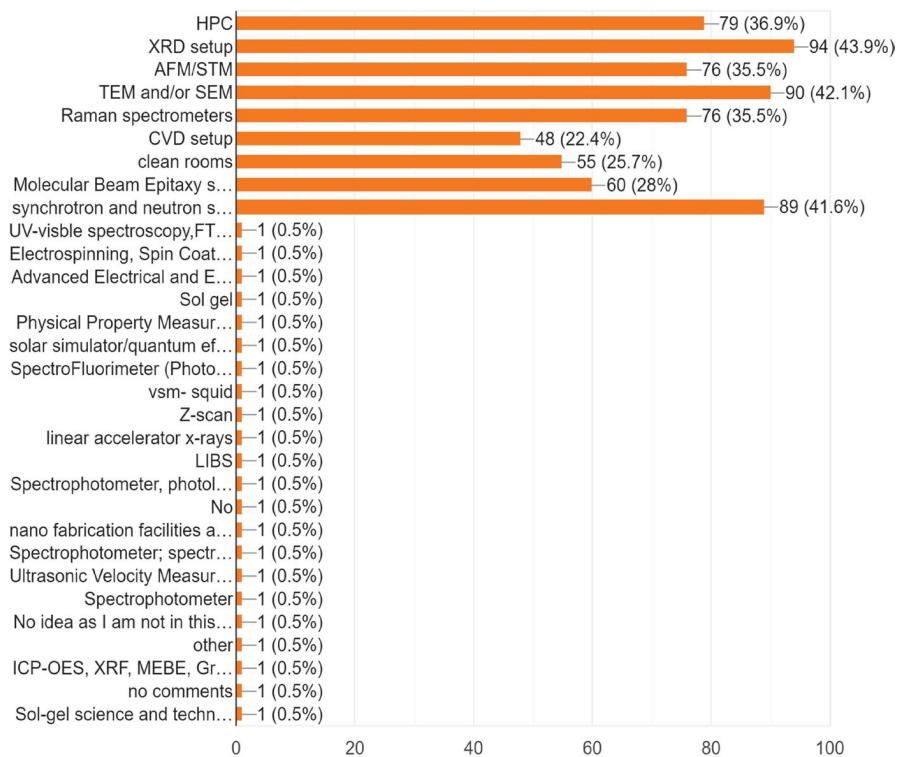

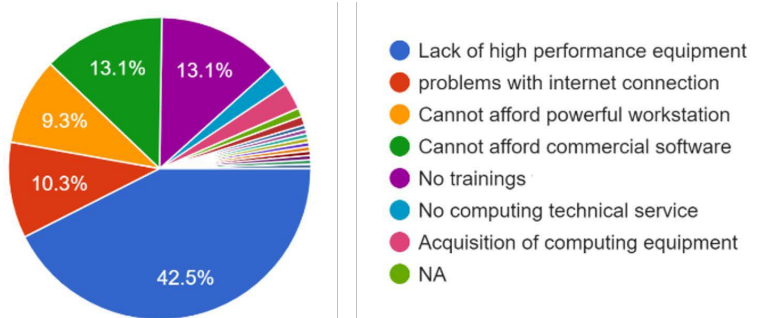

**Figure** 6: Selected responses to the survey launched by the Condensed Matter and Materials Physics WG on the requirements of numerical calculations. Data are updated on June 28th 2022 [12].

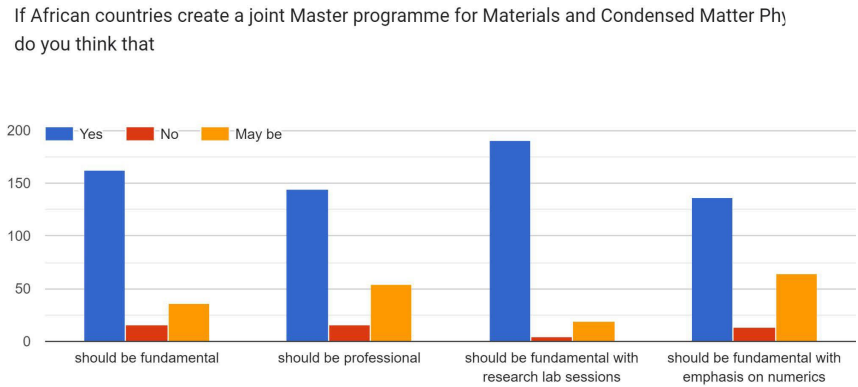

**Figure 7**: Items from the responses to the survey launched by the Condensed Matter and Materials Physics WG. Data are updated on June 28th 2022 [12].

Concerning the LoIs submitted to the ASFAP WGs, 13 out of 64 were addressed to the CM & MP WG. Several ideas were proposed to promote the CM Physics in Africa, and in particular:
- to prioritize research and development of new materials by establishing more centers of excellence - with financial support from the African Union (AU);
- to build a road map to promote teaching and training in CM and MP and to create an inductive environment aiming to engage interested young researchers;
- to minimize the gap between academic and industrial sectors;
- to initiate exchange mutual programmes of collaboration between African universities and to advocate for more female researchers in the field.

## 5. Earth Sciences WG

The Earth Science (ES) WG [13] did not receive yet LoIs. To overcome this issue, this WG will call for LoIs under the umbrella of organizations which will be asked to disseminate the call through their existing networks and mailing lists. The ES WG will organize a mini-symposium planned for direct engagement with the African Earth Sciences community. The event will include an introduction to ASFAP and a keynote conference. During the event, the participants will discuss the content of the questionnaire which will be addressed to involved African researchers in Earth Sciences.

**Outcomes and remarks:**

The essential conclusions of the presentations and the discussion sessions of the above-mentioned WGs can be summarized as follows into two categories.

- The first one concerns a modified approach that is essentially related to the working groups:
- reconsidering a new structuring of the WGs possibly through merging some of them, and to assist creating a constructive synergy by sharing surveys' results, common events, meetings, as well as, data;

- nominating some active focal points in different African countries in order to facilitate targeting various official channels aiming at attracting science and education Ministries, scientific organizations and institutes, in addition to the potential national and international science societies;

- boosting gender balance since women are still underrepresented in the participant communities of ASFAP activities.

- The second category of the concluding remarks deals with the recommendations, which can be summarized as follows:
- Africa needs to define the priorities for regional centers;
- ASFAP needs to keep interaction with different scientific institutions and organization such as ICTP (Italy) and EAIFR (Rwanda) and other regional centers;
- there is a large consensus of the necessity to join forces to build a virtual university (and for other disciplines), and to set-up regional Centers for example for instrumentation and detectors (from small to big equipment).